\documentclass[graybox,graphicx]{svmult}

\usepackage{mathptmx}       
\usepackage{helvet}         
\usepackage{courier}        
\usepackage{type1cm}        
\usepackage{makeidx}         
\usepackage{graphicx}        
\usepackage{multicol}        
\usepackage[bottom]{footmisc}

\makeindex             

\begin{document}

\title*{Modeling neutrino-nucleus interactions for neutrino oscillation experiments}
\author{G. D. Megias, S. Dolan and S. Bolognesi}
\institute{G. D. Megias \at Departamento de F\'isica At\'omica, Molecular y Nuclear,
Universidad de Sevilla, 41080 Sevilla, Spain and DPhP, IRFU, CEA Saclay, 91191 Gif-sur-Yvette, FRANCE \email{megias@us.es}
\and S. Dolan \at IN2P3-CNRS, Laboratoire Leprince-Ringuet, Palaiseau 91120 and DPhP, IRFU, CEA Saclay, 91191 Gif-sur-Yvette, FRANCE \email{stephen.dolan@llr.in2p3.fr} \and S. Bolognesi \at DPhP, IRFU, CEA Saclay, 91191 Gif-sur-Yvette, FRANCE \email{sara.bolognesi@cea.fr}}
%
%
\maketitle
\vspace*{-3.05cm}
\abstract{We present our recent progress on the relativistic modeling of neutrino-nucleus reactions for their implementation in MonteCarlo event generators (GENIE,
NEUT) employed in neutrino oscillation experiments. We compare charged-current neutrino ($\nu$) and antineutrino ($\bar\nu$) cross sections obtained within the SuSAv2 model, which is based on the Relativistic Mean Field theory and on the analysis of the superscaling behavior exhibited by ($e,e'$) data. We
also evaluate and discuss the impact of multi-nucleon excitations arising from 2p-2h states excited by the action
of weak forces in a fully relativistic framework, showing for the first time their implementation in GENIE and their comparison with recent T2K data.}

\vspace{0.24cm}

\begin{figure}
  \begin{minipage}{\textwidth}
    \begin{flushleft}
\includegraphics[width=4cm, angle=270]{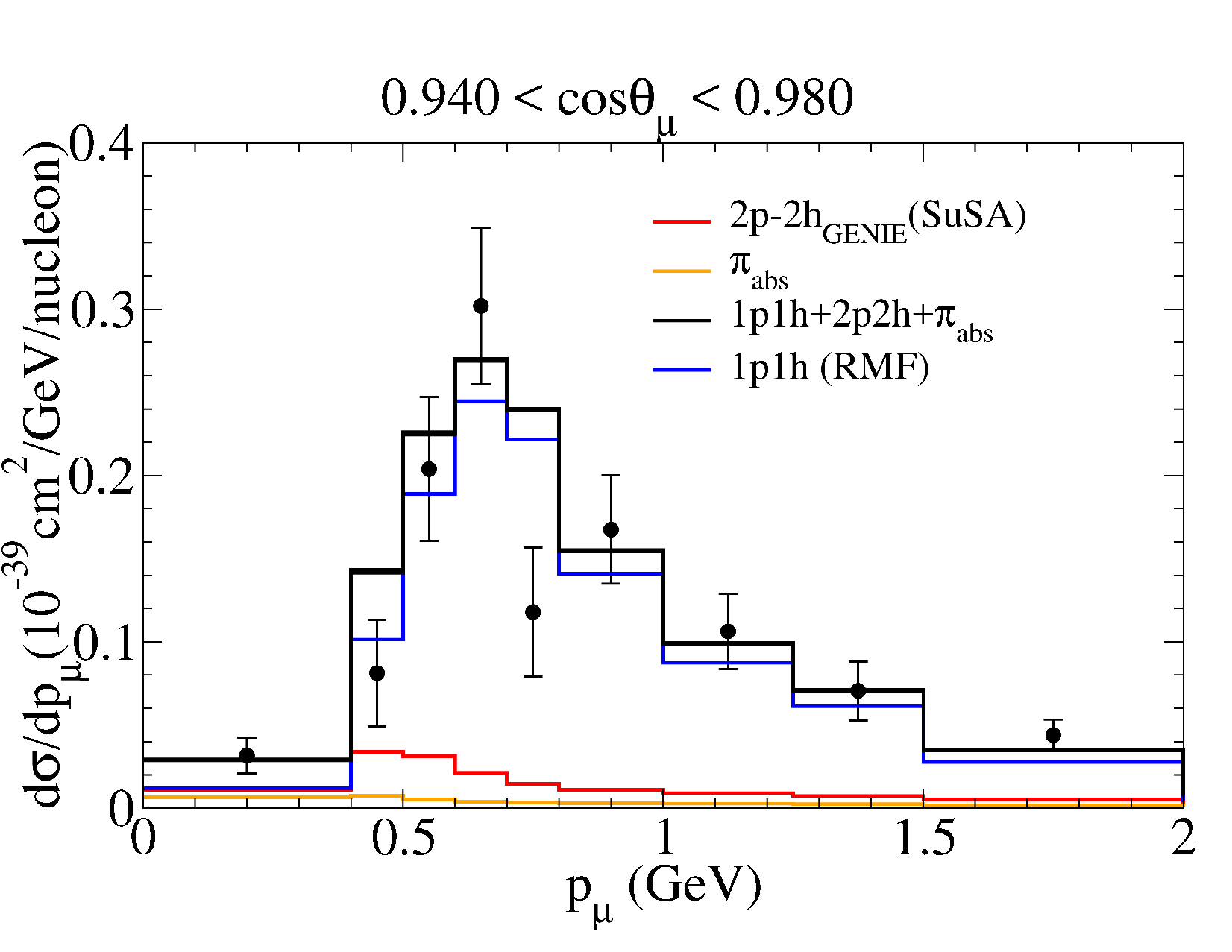} \includegraphics[width=4cm, angle=270]{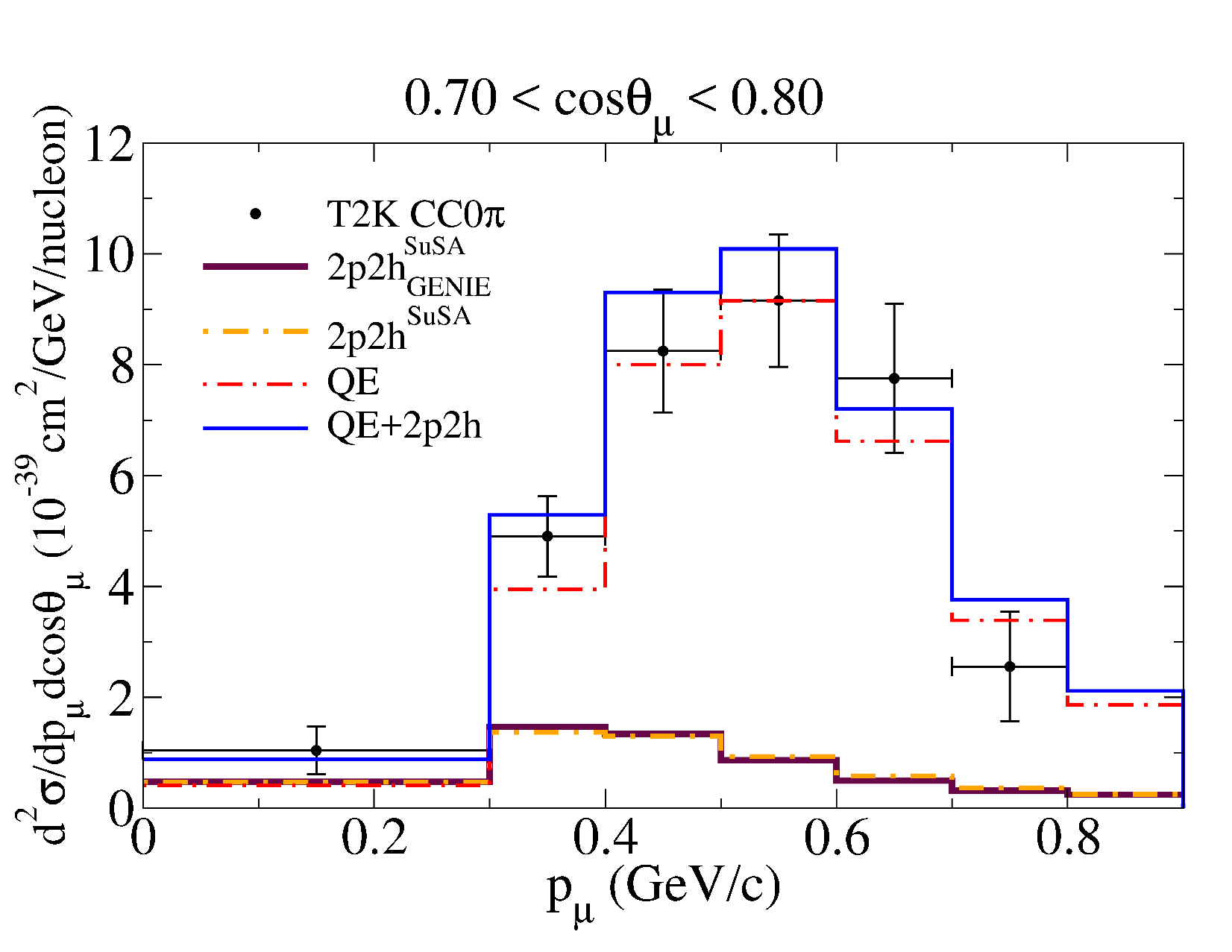}
\end{flushleft}\vspace{-0.54cm}

\caption{(Color online) Left panel: Comparison of T2K CC0$\pi$Np data on $^{12}$C~\cite{cc0pinp} for 0 protons above 500 MeV/c with the SuSAv2-MEC model and the 2p2h GENIE implementation (2p2h$_{\mbox{\tiny GENIE}}^{\mbox{\tiny SuSA}}$). Pion absorption effects are also included. Right panel: Comparison of T2K CC0$\pi$ data on $^{12}$C~\cite{cc0pi} with the SuSAv2-MEC model (QE+2p2h). Comparison between 2p2h GENIE implementation (2p2h$_{\mbox{\tiny GENIE}}^{\mbox{\tiny SuSA}}$) and the microscopic calculation (2p2h$^{\mbox{\tiny SuSA}}$) is also shown.
\label{fig:fig0} }\vspace{-0.49cm}

\end{minipage}
\end{figure}

Current efforts in long-baseline $\nu$ experiments are aimed at improving knowledge of $\nu$ oscillations, where the development and implementation of realistic $\nu$-nucleus interaction models are essential to constrain experimental uncertainties. The current state of the art for experimental systematics is in the region of 5-10$\%$~\cite{Abe2016} and are mostly related to flux and cross section predictions (3-4$\%$). A decrease of 2-3$\%$ on these uncertainties would allow to shorten running time and experimental costs (reducing by half either the experimental exposure or the detector volume) while increasing the sensitivity to determine $\nu$ mass hierarchy or CP violation in the neutrino sector. Such a reduction of systematics will therefore represent an essential step toward understanding the matter-antimatter asymmetry in the Universe, whilst also aiding in other areas of fundamental physics, such as the analysis of supernovae explosions and the search for both sterile $\nu$ and proton decay. Accordingly, an accurate understanding of $\nu$ interaction physics is essential for current and upcoming experiments. Thus, the SuSAv2-MEC approach~\cite{Megias:2016ee,Megias:2016nu,Ivanov14} is applied to the analysis of data from $\nu$ oscillation experiments with the aim of shedding light on the systematics arising from nuclear effects in both initial and final states. For practical purposes, the SuSAv2 model and the 2p-2h MEC contributions can be described in a simple way for different kinematics and nuclei~\cite{MegiasOxygen,MegiasMinerva,ARFG,nppairs,density}, translating sophisticated and demanding microscopic calculations into a relatively straightforward formalism hence easing its implementation in event generators. In Fig.~\ref{fig:fig0}, we show the comparison of the SuSAv2-MEC model, which is based on relativistic, microscopic calculations~\cite{simomec,megiasmec}, with T2K CC0$\pi$Np data~\cite{cc0pinp,dolan} for 0 protons above 500 MeV/c (left panel). The 1p1h channel corresponds to RMF-based calculations and the effect of $\pi$ emission followed by re-absorption in the nuclear medium is provided by the GENIE $\nu$-nucleus event generator~\cite{genie}. The 2p2h channel is generated for the first time by new implementation of the SuSAv2-MEC model within GENIE. The accurate modeling of 2p2h microscopic calculations (thick dot-dashed lines) within GENIE (solid maroon line) can be observed in the right panel, where a comparison of the full SuSAv2-MEC model with CC0$\pi$ data~\cite{cc0pi} is also shown. Its capability to describe data in a wide energy range and its ease to be implemented in event generators makes the SuSAv2-MEC model a promising candidate to reduce experimental systematics in current and future $\nu$ experiments.\vspace{-0.68cm}

%

\vspace{-0.49cm}

\begin{acknowledgement}
This work was partially supported by the Spanish Ministerio
de Economia y Competitividad and ERDF (European
Regional Development Fund) under contracts
FIS2017-88410-P, and by the Junta de Andalucia (grant No.
FQM160). GDM acknowledges support from a Junta de Andalucia fellowship (FQM7632, Proyectos de Excelencia
2011). We acknowledge the support of CEA, CNRS/IN2P3 and
P2IO, France; and the MSCA-RISE project
JENNIFER, funded by EU grant n.644294, for supporting
the EU-Japan researchers’ mobility.
\end{acknowledgement}\vspace*{-0.95cm}
%
%
%

\vspace*{-0.5cm}
\end{document}